\documentclass[showpacs,pra]{revtex4}%
\usepackage{graphicx}
\usepackage{amsmath}
\usepackage{amsfonts}
\usepackage{amssymb}%
\begin{document}

\title{Optimal Co-linear Gaussian Beams for Spontaneous Parametric Down-Conversion}
\author{Ryan S. Bennink}
\affiliation{Oak Ridge National Laboratory, Oak Ridge, Tennessee 37831, USA}
\email{benninkrs@ornl.gov}
\date{\today}

\begin{abstract}
I investigate the properties of spontaneous parametric down-conversion (SPDC)
involving co-linear Gaussian spatial modes for the pump and the photon
collection optics. Approximate analytical and numerical results are obtained
for the peak spectral density, photon bandwidth, pair collection probability,
heralding ratio, and spectral purity, as a function of crystal length and beam
focusing parameters. I address the optimization of these properties
individually as well as jointly, and find focusing conditions that
simultaneously bring the pair collection probability, heralding ratio, and
spectral purity to near-optimal values. These properties are also found to
be nearly scale invariant, that is, ultimately independent of crystal length.
The results obtained here are expected to be useful for designing SPDC
sources with high performance in multiple categories for the next generation
of SPDC applications.
\end{abstract}

\maketitle

\section{Introduction}\label{Intro Sxn}

Spontaneous parametric down-conversion (SPDC) is a process
widely used to produce both entangled photon pairs and heralded single
photons. SPDC photon sources play a key role in quantum cryptography, optical
quantum computing, quantum metrology, and in fundamental studies of quantum
mechanics. In many applications, a nonlinear optical crystal is pumped by a
focused laser beam and the emitted SPDC light is collected into a pair of
optical waveguides, such as an integrated optical circuit \cite{Politi2009_IEEE} or a
fiber network \cite{SECOQC2009}. Efforts over the past decade have led to
some very bright SPDC sources \cite{Wong2006,Fedrizzi2007}. As brightness
has improved, interest has progressed from simply making SPDC sources brighter
to controlling other properties of the emitted photons, such as the spectral
entanglement in each photon pair \cite{Grice2001,U'Ren2007,Mosley2009,Humble2007}.
 While several papers in recent years \cite{Kurtsiefer2001, Bovino2003, Dragan2004, Andrews2004, Castelletto2005, Ljunggren2005, Ling2008, Kolenderski2009} have addressed
the question of how to focus the pump and/or collection optics optimally, some
important questions remain. Since different studies have invoked slightly
different assumptions and optimized slightly different measures, it is not
clear how pump and/or collection focusing simultaneously affects all the quantities
that are generally of interest, and what trade-offs (if any) exist between
these quantities. Furthermore, the scaling laws for the optimized quantities
and optimal parameter values are not readily apparent (e.g. whether it is
possible to halve the crystal length and, through a change of beam parameters,
obtain similar performance).

To address such questions, I present here a new study of SPDC for the case in
which the pump and collecting optics define co-linear Gaussian spatial modes.
 While the theory is general, the envisioned context is SPDC in a
periodically poled nonlinear crystal with emission in the visible or
telecommunication spectral range (a common and useful configuration). In
this study I consider five properties of the collected biphoton state that are
commonly of interest: the joint spectral density; the photon bandwidths; the
pair collection probability; the heralding ratio (pair/single photon
collection ratio); and the spectral entanglement. These properties are
calculated---in some cases analytically, in some cases numerically---as
functions of experimental parameters, yielding predictions for the absolute
values of the properties as well as for the parameter values that optimize
each property. Additionally, this study reveals several scaling laws and
shows that some properties can be jointly optimized while others require a
trade-off. Many of the results presented here are new, with a few appearing
to differ from predictions by others.

The paper is organized as follows: Section \ref{Theory Sxn} establishes the
foundational equations describing Gaussian optical modes and the quantum
physics of SPDC. Sections \ref{Intensity Sxn}-\ref{Herald Sxn} derive
expressions for the five properties mentioned above and discuss their
dependence on experimental parameters. Section \ref{Discussion Sxn}
discusses the results of this study in light of prior works, and Section
\ref{Summary Sxn} summarizes the main conclusions.

\section{SPDC with Gaussian Spatial Modes}\label{Theory Sxn}

SPDC is the lowest-order effect of parametric interaction
between a strong pump (p) field and two other fields, designated as signal (s)
and idler (i), initially in the vacuum state. The quantum Hamiltonian
governing this interaction is
\begin{equation}
\hat{H}=\int\varepsilon_{0}\left(  \boldsymbol{\chi}^{(2)}(\mathbf{r}%
):\mathbf{\hat{E}}^{+}(\mathbf{r},t)\mathbf{\hat{E}}^{-}(\mathbf{r}%
,t)\mathbf{\hat{E}}^{-}(\mathbf{r},t)+\text{H.c.}\right)  \,d^{3}%
\mathbf{r}\label{Hamiltonian, general}%
\end{equation}
where $\mathbf{r}=(x,y,z)$ is the spatial coordinate, $t$ is time,
$\boldsymbol{\chi}^{(2)}(\mathbf{r})$ is the nonlinear susceptibility tensor, and
$\mathbf{\hat{E}}^{+}\mathbf{(r},t)=\mathbf{\hat{E}}^{-}(\mathbf{r}%
,t)^{\dagger}$ is the positive-frequency part of the electric field quantum
operator. In canonical treatments, the field is expanded in terms of an
orthonormal set of modes. Since the interest here is in SPDC involving
Gaussian spatial modes, I choose a modal expansion containing a Gaussian mode
for each frequency $\omega$. It will be sufficient to consider just these
Gaussian modes until Section \ref{Herald Sxn}. In this case $\mathbf{\hat{E}%
}^{+}(\mathbf{r},t)$ may be written as
\begin{equation}
\mathbf{\hat{E}}^{+}(\mathbf{r},t)=\int_{0}^{\infty}d\omega\sqrt{\frac
{\hbar\omega}{4\pi c \varepsilon_{0}}}\mathbf{E}_{\omega}(\mathbf{r})e^{-i\omega
t}\hat{a}_{\omega}(t)+\text{irrelevant non-Gaussian modes}\label{E modal expansion}%
\end{equation}
where $\mathbf{E}_{\omega}(\mathbf{r})$ is the electric field function for the
(Gaussian) mode of frequency $\omega$ and $\hat{a}_{\omega}^{\dagger}(t)$ is the
operator that creates a photon in that mode.

The state resulting from the interaction (\ref{Hamiltonian, general}) is
obtained by applying the operator $\exp\left[  -\frac{i}{\hbar}\int_{-\infty
}^{\infty}dt\,\hat{H}(t)\right]  $ to the initial state. The part of the state
corresponding to SPDC, that is, the creation of a single pair of photons, is
just the first-order term:
\begin{equation}
|\Psi_{\text{SPDC}}\rangle=-\frac{i}{\hbar}\int_{-\infty}^{\infty}dt\,\hat
{H}(t)\,|\text{initial}\rangle.\label{SPDC state, general}%
\end{equation}
Putting (\ref{Hamiltonian, general}) and (\ref{E modal expansion}) into
(\ref{SPDC state, general}) gives
\begin{equation}
|\Psi_{\text{SPDC}}\rangle=-i\int_{0}^{\infty}d\omega_{\text{s}}%
d\omega_{\text{i}}\,\psi(\omega_{\text{s}},\omega_{\text{i}})\hat{a}%
_{\omega_{\text{s}}}^{\dagger}\hat{a}_{\omega_{\text{i}}}^{\dagger}%
|\text{vac}\rangle
\end{equation}
where $|$vac$\rangle$ is the vacuum state of the signal and idler and
\begin{equation}
\psi(\omega_{\text{s}},\omega_{\text{i}})=\sqrt{\frac{2\pi^{2}\hbar
N_{\text{p}}}{\varepsilon_{0}\lambda_{\text{p}}\lambda_{\text{s}}%
\lambda_{\text{i}}}}s(\omega_{\text{p}})\mathcal{O}(\omega_{\text{s}}%
,\omega_{\text{i}})\label{amplitude def}%
\end{equation}
is the SPDC amplitude. Here $\varepsilon_{0}$ is the vacuum permittivity,
$\lambda_{j}=2\pi c/\omega_{j}$ is the free space wavelength of field $j$
($j=$ p,s,i), and $\omega_{\text{p}}=\omega_{\text{s}}+\omega_{\text{i}}$. The
pump field has been assumed to be in a coherent state with independent
spectral and spatial dependence. Accordingly, the operator $\hat{a}%
_{\omega_{\text{p}}}$ has been replaced by the complex amplitude
$a(\omega_{\text{p}})=s(\omega_{\text{p}})\sqrt{N_{_{\text{p}}}},$ where
$s(\omega_{\text{p}})$ is the pump spectral amplitude (with $\int\left|
s(\omega)\right|  ^{2}\,d\omega=1$) and $N_{\text{p}}$ is the mean number of
pump photons (the pump energy divided by $\hbar\omega_{\text{p}}$). The
quantity
\begin{equation}
\mathcal{O}(\omega_{\text{s}},\omega_{\text{i}})\equiv\int
\limits_{\text{medium}}\boldsymbol{\chi}^{(2)}(\mathbf{r}):\mathbf{E}%
_{\omega_{\text{p}}}(\mathbf{r})\mathbf{E}_{\omega_{\text{s}}}^{\ast
}(\mathbf{r})\mathbf{E}_{\omega_{\text{s}}}^{\ast}(\mathbf{r})\mathbf{\,}%
d^{3}\mathbf{r}\label{overlap def}%
\end{equation}
is the spatial overlap of the pump, signal, and idler modes in the medium,
which generalizes the $\operatorname{sinc}$ phase matching function
encountered in plane-wave treatments of SPDC \cite{Hong1985_PRA}. The medium
is taken to be a bulk material of length $L$ centered at the origin, with
cross section large enough to contain the mode functions. The material may
be ferroelectrically poled so that $\boldsymbol{\chi}^{(2)}(\mathbf{r})$
alternates sign with spatial period $\Lambda$.

The properties of the SPDC state can be calculated once particular forms are
chosen for the pump spectrum $s(\omega_{\text{p}})$ and the mode functions
$\mathbf{E}_{\omega_{j}}(\mathbf{r})$. Consideration will be restricted to
modes of the form
\begin{equation}
\mathbf{E}(\mathbf{r})=\frac{\mathbf{e}}{\sqrt{\pi/2}}\frac{w}{q}\exp\left[
-\frac{x^{2}+y^{2}}{q}+ikz\right] \label{Gaussian mode}%
\end{equation}
which describes a linearly polarized, paraxial Gaussian beam with waist at the
origin, propagating along the $z$ axis in an optically uniform medium. Here
$w$ is the waist size, $\mathbf{e}$ is the polarization unit vector, $k =
n\omega/c$ is the wavenumber, $n$ is the refractive index, and $q=w^{2}%
+2iz/k$. This choice is in part motivated by the fact that some of the best
SPDC sources in existence involve Gaussian beams co-propagating along one of
the principle refractive axes of a transparent crystalline material, such as
potassium titanyl phosphate or lithium niobate. (The
anisotropy of the refractive index may be safely ignored in such cases.)
 Additionally, symmetry indicates that the spatial overlap is largest when
the modes co-propagate and have their waists co-located at the center of the
crystal. This leaves the waist sizes $w_{\text{p}},w_{\text{s}},w_{\text{i}%
}$ as the free parameters for mode optimization.

Using modes of the form (\ref{Gaussian mode}), the mode overlap (\ref{overlap
def}) is
\begin{equation}
\mathcal{O}=\frac{\sqrt{\epsilon}\chi_{\text{eff}}^{(2)}}{\left(
\pi/2\right)  ^{3/2}}\int\frac{w_{\text{p}}w_{\text{s}}w_{\text{i}}%
}{q_{\text{p}}q_{\text{s}}^{\ast}q_{\text{i}}^{\ast}}\exp\left[  -\left(
x^{2}+y^{2}\right)  \left(  \frac{1}{q_{\text{p}}}+\frac{1}{q_{\text{s}}%
^{\ast}}+\frac{1}{q_{\text{i}}^{\ast}}\right)  +i(\Delta k+mK)z\right]
\mathbf{\,}dx\,dy\,dz.
\end{equation}
Here $\Delta k=k_{\text{p}}-k_{\text{s}}-k_{\text{i}}$ is the wavenumber
mismatch, $K=2\pi/\Lambda$ is the poling spatial frequency, $m$ is the order of
quasi-phase matching, $\chi_{\text{eff}}^{(2)}\equiv\boldsymbol{\chi}%
^{(2)}:\mathbf{e}_{\text{p}}\mathbf{e}_{\text{s}}\mathbf{e}_{\text{i}}$ is the
effective nonlinear coefficient, and $\epsilon$ is an efficiency factor that
includes Fresnel loss and the Fourier coefficient of the $m$th harmonic of the poling
spatial function. Integrating over $x$ and $y$ yields
\begin{equation}
\mathcal{O}=\sqrt{\frac{8\epsilon}{\pi}}\chi_{\text{eff}}^{(2)}w_{\text{p}%
}w_{\text{s}}w_{\text{i}}\int_{-L/2}^{L/2}\frac{\exp\left[  i(\Delta
k+mK)z\right]  }{q_{\text{s}}^{\ast}q_{\text{i}}^{\ast}+q_{\text{p}%
}q_{\text{i}}^{\ast}+q_{\text{p}}q_{\text{s}}^{\ast}}%
\,dz.\label{Gaussian overlap}%
\end{equation}
Before proceeding, it will be convenient to rewrite (\ref{Gaussian overlap})
in terms of dimensionless quantities. The primary independent variables are
the phase mismatch
\begin{equation}
\Phi\equiv(\Delta k+mK)L\label{phi def}%
\end{equation}
and the focal parameters
\begin{equation}
\xi_{j}\equiv\frac{L}{k_{j}w_{j}^{2}}\label{xi def}%
\end{equation}
where $\xi_{j}\gg1$ ($\ll1$) means that field $j$ ($j=\,$p,s,i ) is focused
strongly (weakly) relative to the length of the crystal.

I also define the auxiliary quantities
\begin{align}
A_{\pm}  & \equiv1+\frac{k_{\text{s}}}{k_{\text{p}}}\frac{\xi_{\text{s}}}%
{\xi_{\text{p}}}\pm\frac{k_{\text{i}}}{k_{\text{p}}}\frac{\xi_{\text{i}}}%
{\xi_{\text{p}}},\label{A def}\\
B_{\pm}  & \equiv\left(  1-\frac{\Delta k}{k_{\text{p}}}\right)  \left(
1+\frac{k_{\text{s}}+\Delta k}{k_{\text{p}}-\Delta k}\frac{\xi_{\text{p}}}%
{\xi_{\text{s}}}\pm\frac{k_{\text{i}}+\Delta k}{k_{\text{p}}-\Delta k}%
\frac{\xi_{\text{p}}}{\xi_{\text{i}}}\right)  ,\label{B def}\\
C  & \equiv\frac{\Delta k}{k_{\text{p}}}\frac{\xi_{\text{p}}^{2}}%
{\xi_{\text{s}}\xi_{\text{i}}}\frac{A_{+}}{B_{+}^{2}},\label{C def}%
\end{align}
and the aggregate focal parameter
\begin{equation}
\xi\equiv\frac{B_{+}}{A_{+}}\frac{\xi_{\text{s}}\xi_{\text{i}}}{\xi_{\text{p}%
}}.\label{xi aggregate def}%
\end{equation}
The quantities $A_{+}$, $B_{+},$ and $\xi$ are independent and uniquely
determine $\xi_{\text{s}}$, $\xi_{\text{i}}$, and $\xi_{\text{p}}$. $C$ is
determined by $A_{+}$ and $B_{+}$. Note that $A_{\pm}$, $B_{\pm}$, and $C$ do
not depend on the absolute values of the focal parameters, but on the focus of
the signal and idler \emph{relative} to the pump. In terms of these
dimensionless quantities, eq. (\ref{amplitude def}) may be written as
\begin{equation}
\psi(\omega_{\text{s}},\omega_{\text{i}})=\sqrt{\frac{8\pi^{2}\epsilon\hbar
n_{\text{s}}n_{\text{i}}N_{\text{p}}L}{\varepsilon_{0}n_{\text{p}}}}\frac
{\chi_{\text{eff}}^{(2)}}{\lambda_{\text{s}}\lambda_{\text{i}}}\frac
{s(\omega_{\text{p}})}{\sqrt{A_{+}B_{+}}}\int_{-1}^{1}\frac{\sqrt{\xi}%
\exp\left[  i\Phi l/2\right]  }{1-i\xi l-C\xi^{2}l^{2}}%
\,dl.\label{amplitude formula}%
\end{equation}
 Eq.\ (\ref{amplitude formula}) will be the starting point for analysis in
sections \ref{Intensity Sxn}-\ref{Purity Sxn}.

Four approximations will be invoked throughout this work in order to simplify
analysis and yield more useful results. These approximations are reasonable
for typical bulk SPDC sources, which may be defined to have the following
characteristics: the length of the medium is $\gtrsim 1 \operatorname{mm}$
 and its refractive index is $\gtrsim 1.5$; the parameteric interaction is
quasi-phase matched with a first-order grating of period $\Lambda \gtrsim 5 \operatorname{\mu m}$;
 and emission is in the visible or telecommunication spectral range, with
$\lambda_{\text{s}} \lesssim 1.6 \operatorname{\mu m}$ and $\lambda_{\text{p}} \lesssim 0.8
\operatorname{\mu m}$. When assessing the accuracy of approximate formulas, these values will be
considered to represent the worst typical case. I will also consider a
``reference source'' consisting of degenerate type II SPDC in $10 \operatorname{mm}$ 
periodically poled potassiam titanyl phosphyate (PPKTP) with a $750 \operatorname{nm}$ pump.
This is similar to several sources that have been demonstrated to
have good performance \cite{Wong2006,Fedrizzi2007,ORNL_source}.

One approximation that will be used liberally is
\begin{equation}
1\pm\frac{\Delta k}{k_{j}}\approx1.\label{delta k small approx}%
\end{equation}
This approximation is motivated by the fact that efficient SPDC occurs when
$\Delta k\approx-mK,$ where $mK/k_{j}$ is typically much smaller than 1.

Another approximation that will prove convenient is
\begin{equation}
C\approx0.\label{C small approx}%
\end{equation}
The actual value of $C$ depends in a complicated way upon the experimental
parameters, but can be shown to obey the bound
\begin{equation}
\left|  C\right|  \leq\frac{\left|  \Delta k\right|  k_{\text{p}}%
}{4k_{\text{s}}k_{\text{i}}}=\left|  \frac{\Phi k_{\text{p}}}{4k_{\text{s}%
}k_{\text{i}}L}-\frac{mKk_{\text{p}}}{4k_{\text{s}}k_{\text{i}}}\right|
\label{C bound}%
\end{equation}
when (\ref{delta k small approx}) is valid. One then has $|C| \lesssim 0.1$ near
phase matching conditions, which turns out to be small enough to make (\ref{C
small approx}) a fair approximation over the range of focusing that is
desirable with respect to the five properties of the state considered here.

The third approximation is that the frequency dependence of (\ref{amplitude
formula}) is determined essentially by the pump spectrum $s(\omega_{\text{p}%
})$ and the frequency dependence of the wave mismatch $\Delta k$. That is, I
take
\begin{equation}
A_{\pm},B_{\pm},\xi_{j},\sqrt{\frac{\epsilon n_{\text{s}}n_{\text{i}}%
}{n_{\text{p}}}}\frac{\chi_{\text{eff}}^{(2)}}{\lambda_{\text{s}}%
\lambda_{\text{i}}}\approx\text{constant}\label{frequency independence approx}%
\end{equation}
over the range of $(\omega_{\text{s}},\omega_{\text{i}})$ for which the
amplitude is appreciable. For (\ref{frequency independence approx}) to hold,
the bandwidths of the photons must be much smaller than an optical frequency.
 Type II and non-degenerate type I SPDC sources typically satisfy this
condition, but frequency-degenerate type I sources and sources made of very short crystals,
which tend to have large bandwidths, may not.

Finally, after section \ref{Intensity Sxn} it will be assumed the wave
mismatch $\Delta k$ has a predominantly linear frequency dependence:
\begin{equation}
\delta k_{j}\approx\frac{n_{j}^{\prime}}{c}\delta\omega_{j}%
\label{linear dispersion approx}%
\end{equation}
where $n_{j}^{\prime}\equiv c\,\partial k_{j}/\partial\omega$ is the group
index of mode $j$ and $\delta\omega_{j}$ $(\delta k_{j})$ denotes a shift from
the nominal frequency (wavenumber) of mode $j$. Again, this approximation is
generally valid for type II and non-degenerate type I SPDC sources, but not
for frequency-degenerate type I sources.

The impact of these approximations, particularly (\ref{C small approx}), will
be addressed more fully in the context of each major result.

\section{Joint Spectral Density}\label{Intensity Sxn}

The joint spectral density $\left|  \psi(\omega
_{\text{s}},\omega_{\text{i}})\right| ^{2}$ is the expected number of photons
pairs, per signal bandwidth per idler bandwidth, emitted into the Gaussian collection
modes. If the collected photons pass through spectral filters of narrow
bandwidth, the effective brightness of the source is determined by the joint
spectral density at the filter frequencies. Let us now consider the problem
of maximizing this quantity, that is, finding the values of $\xi_{\text{p}%
},\xi_{\text{s}},\xi_{\text{i}},\Phi$ that maximize $\left|  \psi
(\omega_{\text{s}},\omega_{\text{i}})\right|  ^{2}$ given the frequencies
$\omega_{\text{p}},\omega_{\text{s}},\omega_{\text{i}}$ and crystal length
$L$. From (\ref{amplitude formula}), it is apparent that maximization
determines certain values $(A_{+}^{\text{max}},B_{+}^{\text{max}}%
,\xi^{\text{max}},\Phi^{\text{max}})$. The beam parameters $\xi_{\text{p}}%
,\xi_{\text{s}},\xi_{\text{i}}$ are then complicated functions of
$(k_{\text{p}},k_{\text{s}},k_{\text{i}},A_{+}^{\text{max}},B_{+}^{\text{max}%
},\xi^{\text{max}},\Phi^{\text{max}})$ via relations (\ref{phi def})-(\ref{xi
aggregate def}). However, substantially more informative results can be
obtained if the $l^{2}$ term in (\ref{amplitude formula}) is neglected. With
$C\approx0$ (approximation (\ref{C small approx})), the function to be
maximized is
\begin{equation}\label{amplitude dim'less factor}
\frac{1}{\sqrt{A_{+}B_{+}}}\times\int_{-1}^{1}\frac{\sqrt{\xi}\exp\left[
i\Phi l/2\right]  }{1-i\xi l}\,dl.
\end{equation}
The two factors are independent and can be maximized separately. $A_{+}%
B_{+}$ can be written as
\[
A_{+}B_{+}=\left(  1-\frac{\Delta k}{k_{\text{p}}}\right)  \left(
1+X_{\text{s}}r_{\text{s}}+X_{\text{i}}r_{\text{i}}\right)  \left(
1+\frac{X_{\text{s}}}{r_{\text{s}}}+\frac{X_{\text{i}}}{r_{\text{i}}}\right)
\]
where
\begin{align}
X_{j}  & \equiv\frac{k_{j}}{k_{\text{p}}}\sqrt{\frac{1+\Delta k/k_{j}%
}{1-\Delta k/k_{\text{p}}},}\\
r_{j}  & \equiv\frac{\xi_{j}}{\xi_{\text{p}}}\sqrt{\frac{1-\Delta
k/k_{\text{p}}}{1+\Delta k/k_{j}}}.
\end{align}
The maximum of $\left(  A_{+}B_{+}\right)  ^{-1/2}$ occurs at $r_{\text{s}%
}=r_{\text{i}}=1$. Invoking approximation (\ref{delta k small approx}) gives
\begin{equation}
\xi_{\text{s}} \approx\xi_{\text{i}} \approx\xi_{\text{p}} \approx \xi \label{equal Rayleigh}%
\end{equation}
and
\begin{equation}
\max\frac{1}{\sqrt{A_{+}B_{+}}}\approx\frac{1}{2}.\label{AB result}%
\end{equation}

\begin{figure}
	\centering
		\includegraphics{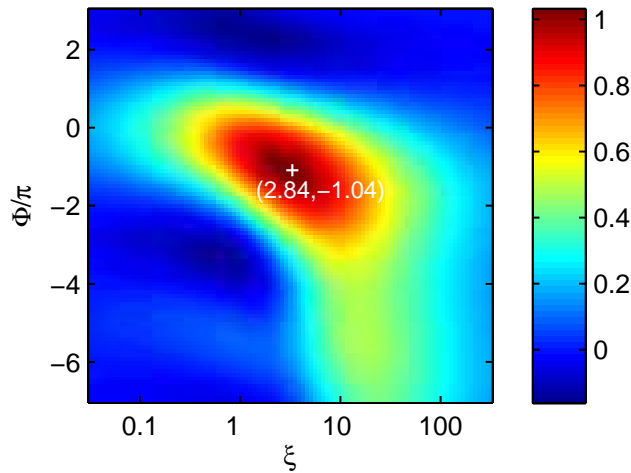}
	\caption{(color online) The spatial overlap factor $F(\xi,\Phi)$ defined in eq.\ (\ref{integral function}). The cross marks the location of the peak.}
	\label{fig: integral function}
\end{figure}

The second factor in (\ref{amplitude dim'less factor}),
\begin{equation}
F(\xi,\Phi)\equiv\int_{-1}^{1}\frac{\sqrt{\xi}\exp\left[  i\Phi l/2\right]
}{1-i\xi l}\,dl\label{integral function}%
\end{equation}
requires numerical evaluation and is plotted in Fig.\ \ref{fig: integral function}. (Note that for $\xi\ll1,$ $F$ is proportional to the usual phase
matching function $\operatorname{sinc}(\Phi/2)$.) The maximum value of $F$ is
$2.06\cdots$ at $\xi=2.84\cdots$, $\Phi=-(1.04\cdots)\pi$. This result,
together with (\ref{AB result}), yields the maximum spectral amplitude
\begin{equation}
\max_{\xi,\Phi}\left|  \psi(\omega_{\text{s}},\omega_{\text{i}})\right|
\approx1.03\sqrt{\frac{8\pi^{2}\hbar\epsilon n_{\text{s}}n_{\text{i}}%
}{\varepsilon_{0}n_{\text{p}}}}\frac{\chi_{\text{eff}}^{(2)}}{\lambda
_{\text{s}}\lambda_{\text{i}}}\sqrt{N_{\text{p}}L}s(\omega_{\text{p}%
})\label{max amplitude}%
\end{equation}
which is obtained under the conditions
\begin{align}
\left.  \xi_{\text{s}}\approx\xi_{\text{i}}\approx\xi_{\text{p}}\right.   &
\approx2.84,\label{cond max amplitude 1}\\
\Phi & \approx-1.04\pi.\label{cond max amplitude 2}%
\end{align}
%\begin{REVISED}
It should be noted that these are precisely the conditions which have long
been known to maximize the efficiency of sum frequency generation and parametric
amplification with Gaussian beams \cite{Boyd1968}. The spectral density $\left|
\psi(\omega_{\text{s}},\omega_{\text{i}})\right|^{2}$ is proportional to the
number of pump photons $N_{\text{p}}$ and to the crystal length.  (In contrast, for the case of
collimated (plane wave) interaction, the spectral density would grow quadratically
with the crystal length.)  It may also be noted that, for all other factors being equal,
short-wavelength sources are brighter than long-wavelength sources.
%\end{REVISED}

\begin{figure}
	\centering
		\includegraphics{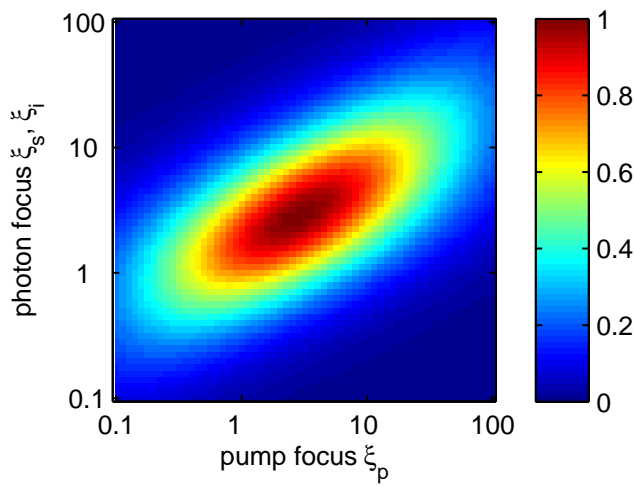}
	\caption{(color online) Peak joint spectral density, normalized to the global maximum, as a function of the pump and photon mode focus.}
	\label{fig: peak density image}
\end{figure}

Since it may not always be possible to implement or verify condition
(\ref{cond max amplitude 1}) accurately, it is worth examining how sensitive
the peak spectral density is to the focal parameters. Specifically, I
compare $\max_{\Phi}\left|  \psi\right|  ^{2},$ which is a function of
$(\xi_{\text{p}},\xi_{\text{s}},\xi_{\text{i}}),$ to the global maximum value
$\max_{\xi,\Phi}\left|  \psi\right|  ^{2}$. I constrain the parameter space by
the condition $\xi_{\text{s}}=\xi_{\text{i}},$ which is not only optimal in
regard to the spectral density, but as will be shown in later sections, is
also optimal or near-optimal for other quantities of interest. Fig.\ \ref{fig: peak density image}
shows the relative brightness $\max_{\Phi}\left|  \psi\right|  ^{2}/\max
_{\xi,\Phi}\left|  \psi\right|  ^{2}$ as a function of $\xi_{\text{p}}$ and
$\xi_{\text{s}}=\xi_{\text{i}}$. It can be seen that the regime of good
focusing is rather broad: the waist size (and correspondingly, the angular
divergence) must be made smaller or larger by roughly a factor of 5 to reduce
the peak spectral density to half its maximum value. Over this range the
optimal phase $\Phi$ (not shown) varies from $-\pi/2$ to $-3\pi/2,$ meaning
that the frequency of the peak shifts slightly with focusing. If the signal
and idler focus are fixed, the optimal pump focus is given by $\xi_{\text{p}%
}=\xi_{\text{s}}(=\xi_{\text{i}})$. However, if the pump focus is fixed, the
optimal signal and idler focus are given by $\xi_{\text{s}}=\xi_{\text{i}%
}\approx\sqrt{2.84\xi_{\text{p}}}$.

These results are robust to approximation: for the worst case defined in
section \ref{Theory Sxn}, the ``optimal'' values given in (\ref{cond max
amplitude 1}) and (\ref{cond max amplitude 2}) differ from the true optimal
values by 5\% and 10\% respectively, but, due to the broadness of $\psi$ as a
function of these parameters, yield an amplitude that is within 1\% of the
actual maximum value.

\section{Photon Bandwidth with Monochromatic Pumping}\label{Bandwidth Sxn}

SPDC photons are sometimes employed in systems with
limited optical bandwidth. In designing sources for such systems it is
helpful to know how the photon bandwidths depend on the design parameters. One
case of particular interest is when the pump bandwidth is chosen to maximize
the spectral purity of the photons. That case will be addressed in section
\ref{Purity Sxn}. Another particularly interesting case is that of
monochromatic pumping, which will be addressed now.

With a monochromatic pump, and under approximation (\ref{frequency
independence approx}), the spectral dependence of (\ref{amplitude formula})
arises primarily from the dispersion of the phase mismatch $\Phi = (\Delta k + mK)L$.
The phase mismatch may be expanded as
Taylor series in $\delta\omega_{\text{s}}$ and $\delta\omega_{\text{i}}$,
the deviation of the signal and idler frequencies from their nominal values.  Assumption
(\ref{linear dispersion approx}) then gives
\begin{equation}
\Phi \approx \Phi_0 + \left(  \frac{n_{\text{p}}^{\prime}}{c}(\delta\omega
_{\text{s}}+\delta\omega_{\text{i}})-\frac{n_{\text{s}}^{\prime}}{c}%
\delta\omega_{\text{s}}-\frac{n_{\text{i}}^{\prime}}{c}\delta\omega_{\text{i}%
}\right) L \label{phase mismatch taylor}%
\end{equation}
where $\Phi_0$ is the phase mismatch at the nominal frequencies. 
Monochromatic pumping constrains the frequencies to $\delta
\omega_{\text{s}}=-\delta\omega_{\text{i}},$ yielding 
\begin{equation}
\Phi \approx \Phi_0 + \frac{(n_{\text{s}}^{\prime}-n_{\text{i}}^{\prime})L}%
{c}\delta\omega_{\text{s}}.\label{phase-frequency relation 1}%
\end{equation}
The fact that $\Phi$ is (approximately) a linear function of $\omega_\text{s}$ means
that the photon bandwidth $\Delta\omega_{\text{s}}$ ($=\Delta\omega_{\text{i}}$) can
be expressed in terms of a dimensionless ``phase mismatch bandwidth'' $\Delta\Phi$,
which is the width of 
%\begin{REVISED}
$|\psi|^2$ when expressed as function of $\Phi$.

\begin{figure}
	\centering
		\includegraphics{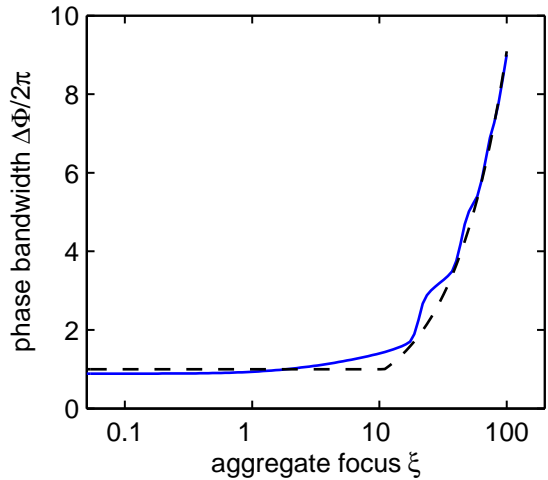}
	\caption{(color online) The normalized photon bandwidth as a function of focusing.  The dashed line is the heuristic formula (\ref{Phi bandwidth heuristic 1}).}
	\label{fig: cw bandwidth}
\end{figure}

Under approximation $C\approx0$, $|\psi|^2$ is proportional to $|F(\xi,\Phi)|^2$.
The full-width at half maximum (FWHM) of $\left| F(\xi,\Phi)\right| ^{2}$
(see Fig.\ \ref{fig: integral function}) was calculated numerically and is plotted as
the solid line in Fig.\ \ref{fig: cw bandwidth}.
%\end{REVISED}
I find that the behavior of $\Delta\Phi$ in this case is captured
well by the heuristic formula
\begin{equation}
\Delta\Phi\sim2\pi\max(1,\xi/10),\label{Phi bandwidth heuristic 1}%
\end{equation}
shown as the dashed line.  Combining (\ref{phase-frequency relation 1}) and (\ref{Phi bandwidth heuristic
1}) yields the photon bandwidths
\begin{equation}
\Delta\omega_{\text{s}}=\Delta\omega_{\text{i}}\sim\frac{2\pi c}{\left|
n_{\text{s}}^{\prime}-n_{\text{i}}^{\prime}\right|  }\max\left(  \frac{1}%
{L},\frac{1}{10b}\right) \label{bandwidth heuristic 1}%
\end{equation}
where $b\equiv L/\xi$ is the aggregate confocal length of the modes (reducing
to $k_{\text{p}}w_{\text{p}}^{2}$ when $\xi_{\text{s}}=\xi_{\text{i}}%
=\xi_{\text{p}}$). When focusing is weak to moderate ($\xi\lesssim10$), the
photon bandwidth is determined by the crystal length (the well-known $1/L$
dependence). 
%\begin{REVISED}
But when the focusing is strong ($\xi\gtrsim10$), the bandwidth
is larger \cite{Carrasco2006} and determined instead by the confocal length, going as $1/b$.

The reason that tightly focusing the pump increases the bandwidth can be understood as follows:
At negative values of the phase mismatch (i.e. away from the nominal wavelengths),
the photon spatial distribution in the far field takes the form of a ring \cite{Kwiat1995}.  With a collimated pump, a certain change in wavelength makes the ring radius larger than that of the collection modes, and photons of those wavelengths are not collected.  But when the pump is tightly focused, the rings are spatially broadened and the photons partially overlap the collection mode \cite{Zhao2008}.  The larger the spatial broadening, the larger the wavelength change must be for the photon distribution to lie outside the collection mode.   
%\end{REVISED}

Due to the approximation $C=0$, the bandwidth plotted in Fig.\ \ref{fig: cw bandwidth} is not exact.
 For the reference source, the error in $\Delta\omega_{\text{s}}$ is
$\leq3\%$ for $\xi\leq10$ and $\leq10\%$ over the range $10\leq\xi_{\text{p}%
}\leq100$. For the worst typical case, the error is $\leq3\%$ for $\xi\leq1,$
but increases to $10\%$ at $\xi=10$ and varies between $30\%$ and $50\%$ over
the range $10\leq\xi_{\text{p}}\leq100$. Thus bandwidth predictions may not
be highly accurate in the regime of focus-induced spectral broadening.
Nevertheless, even in the worst case the approximate SPDC state $|\tilde{\psi
}\rangle$ (calculated with $C=0$) has a large overlap with the exact state
$|\psi\rangle$ (calculated using (\ref{C bound})): the fidelity $\langle
\tilde{\psi}|\psi\rangle^{2}/\langle\tilde{\psi}|\tilde{\psi}\rangle
\langle\psi|\psi\rangle$ is 0.999 at $\xi=1,$ 0.97 at $\xi=10,$ and 0.91 at
$\xi=100$. Also, the heuristic formula $\Delta\Phi\sim2\pi\max(1,\xi/\xi
_{0})$ remains as accurate as shown in Fig.\ \ref{fig: cw bandwidth} for a suitably chosen value of
$\xi_{0}$.

\section{Pair Collection Probability}\label{Flux Sxn}

When SPDC photons are collected without spectral filtering,
maximizing the pair collection probability
\begin{equation}
P_{\text{si}}=\int\left|  \psi(\omega_{\text{s}},\omega_{\text{i}})\right|
^{2}d\omega_{\text{s}}d\omega_{\text{i}}.\label{flux def}%
\end{equation}
is usually more important than maximizing the peak spectral density.
%\begin{REVISED}
(In most experiments, $P_{\text{si}}$ is proportional to the
coincident photodection rate, or the probability of detecting a pair of photons
following any given pump pulse.)
%\end{REVISED}
Under approximation (\ref{frequency independence approx}), the dominant spectral
dependence arises from $s(\omega_{\text{p}})$ and $\Delta k$. The phase
mismatch (\ref{phase mismatch taylor}) may be written as
\begin{equation}
\Phi \approx \Phi_0 + \left(  \frac{2n_{\text{p}}^{\prime}-(n_{\text{s}}^{\prime
}+n_{\text{i}}^{\prime})}{2c}\delta\omega_{\text{p}}-\frac{n_{\text{s}%
}^{\prime}-n_{\text{i}}^{\prime}}{2c}\delta\omega_{-}\right)  L
\end{equation}
where
$\omega_{-}=\omega_{\text{s}}-\omega_{\text{i}}$. Then $d\omega
_{\text{s}}d\omega_{\text{i}}=(d\omega_{-}d\omega_{\text{p}})/2=d\Phi
d\omega_{\text{p}}\,c/(|n_{\text{s}}^{\prime}-n_{\text{i}}^{\prime}|L)$,
giving
\begin{align}
P_{\text{si}}  & =\frac{c}{\left|  n_{\text{s}}^{\prime}-n_{\text{i}}^{\prime
}\right|  L}\int\left|  \psi\right|  ^{2}\,d\omega_{\text{p}}d\Phi\\
& =\frac{8\pi^{2}\hbar c\epsilon n_{\text{s}}n_{\text{i}}}{\varepsilon
_{0}n_{\text{p}}\left|  n_{\text{s}}^{\prime}-n_{\text{i}}^{\prime}\right|
}\left(  \frac{\chi_{\text{eff}}^{(2)}}{\lambda_{\text{s}}\lambda_{\text{i}}%
}\right)  ^{2}\frac{N_{\text{p}}}{A_{+}B_{+}}\int\left|  s(\omega_{\text{p}%
})\int_{-1}^{1}\frac{\sqrt{\xi}\exp\left[  i\Phi l/2\right]  }{1-i\xi
l-C\xi^{2}l^{2}}\,dl\right|  ^{2}d\omega_{\text{p}}d\Phi.
\end{align}
Under the approximation $C\approx0$, the only place $\Phi$ appears is in the
exponential function. Using $\int\exp\left[  i\Phi(l-l^{\prime})/2\right]
\,d\Phi=4\pi\delta(l-l^{\prime})$ and $\int\left|  s(\omega_{\text{p}%
})\right|  ^{2}d\omega_{\text{p}}=1,$ one obtains
\begin{equation}
P_{\text{si}}\approx\frac{32\pi^{3}\hbar c\epsilon n_{\text{s}}n_{\text{i}}%
}{\varepsilon_{0}n_{\text{p}}\left|  n_{\text{s}}^{\prime}-n_{\text{i}%
}^{\prime}\right|  }\left(  \frac{\chi_{\text{eff}}^{(2)}}{\lambda_{\text{s}%
}\lambda_{\text{i}}}\right)  ^{2}\frac{N_{\text{p}}\xi}{A_{+}B_{+}}\int
_{-1}^{1}\frac{dl}{1+\xi^{2}l^{2}}%
\end{equation}
which may be directly integrated to give
\begin{equation}
P_{\text{si}}\approx\frac{64\pi^{3}\hbar c\epsilon n_{\text{s}}n_{\text{i}}%
}{\varepsilon_{0}n_{\text{p}}\left|  n_{\text{s}}^{\prime}-n_{\text{i}%
}^{\prime}\right|  }\left(  \frac{\chi_{\text{eff}}^{(2)}}{\lambda_{\text{s}%
}\lambda_{\text{i}}}\right)  ^{2}\frac{\arctan\left(  \xi\right)  }{A_{+}%
B_{+}}N_{\text{p}}.\label{coinc flux formula}%
\end{equation}

\begin{figure}
	\centering
		\includegraphics{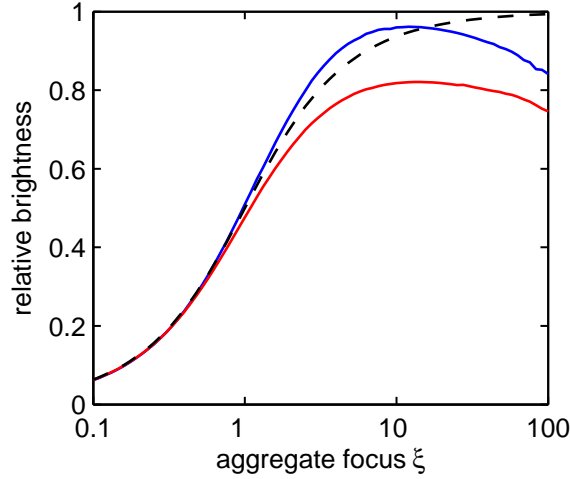}
	\caption{(color online) Dependence of the pair collection probability on the aggregate focus.  The dashed line is eq. (\ref{coinc flux formula}) normalized by its asymptotic maximum.  The solid lines are calculated using ``worst case" values for the parameter $C$ (see discussion surrounding eq.\ (\ref{C bound})) and normalized by the same factor.}
	\label{fig: pair flux line}
\end{figure}

Formula (\ref{coinc flux formula}), plotted as the dashed line in Fig.\ \ref{fig: pair flux line},
suggests that the pair probability can only be optimized in an asymptotic
sense---that there are no finite values of $\xi_{\text{p}},\xi_{\text{s}}%
,\xi_{\text{i}}$ that maximize $P_{\text{si}}$. In reality, the asymptotic
behavior holds only as long as $\xi k_{\text{p}}/(4Lk_{\text{s}}k_{\text{i}%
})\lesssim0.1$. With very tight focusing or very short crystals, the
approximation $C=0$ breaks down, causing $P_{\text{si}}$ to peak near its
asymptotic value (solid lines in Fig.\ \ref{fig: pair flux line}). In the worst typical case the
pair probability error $\left|  1-\langle\tilde{\psi}|\tilde{\psi}%
\rangle/\langle\psi|\psi\rangle\right| $ is $\leq4\%$ for $\xi\leq1,$
$\leq13\%$ for $\xi\leq10,$ and $\leq20\%$ for $\xi\leq100$. For the
reference source described above, the pair probability error is $\leq1.5\%$
for $\xi\leq100$. In any case, $P_{\text{si}}$ evidently has an upper bound
\begin{equation}
P_{\text{si}}\leq\frac{8\pi^{4}\hbar c\epsilon n_{\text{s}}n_{\text{i}}%
}{\varepsilon_{0}n_{\text{p}}\left|  n_{\text{s}}^{\prime}-n_{\text{i}%
}^{\prime}\right|  }\left(  \frac{\chi_{\text{eff}}^{(2)}}{\lambda_{\text{s}%
}\lambda_{\text{i}}}\right)  ^{2}N_{\text{p}}
\label{max pair flux}
\end{equation}
that cannot be exceeded, no matter how long the crystal.
%\begin{REVISED}
(This somewhat surprising result will be discussed in section \ref{Discussion Sxn}.)
%\end{REVISED}
For type II SPDC sources made of
PPKTP or periodically poled lithium niobate (PPLN), eq.\ (\ref{max pair flux}) predicts that brightnesses exceeding $10^{-9}$ collected pairs/pump photon should be achievable---roughly an order of magnitude brighter than the brightest existing sources.

\begin{figure}
	\centering
		\includegraphics{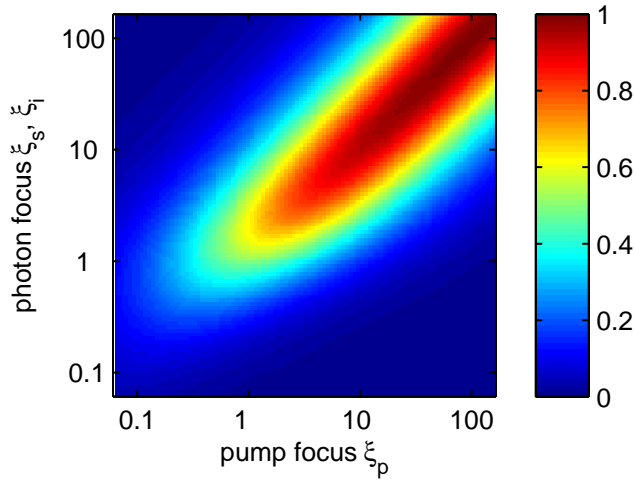}
	\caption{(color online) Dependence of the pair collection probability on the focus of the pump and photon modes.}
	\label{fig: pair flux image}
\end{figure}

The dependence of $P_{\text{si}}$ on $\xi_{\text{p}}$ and $\xi_{\text{s}}%
=\xi_{\text{i}}$ is shown in Fig.\ \ref{fig: pair flux image}. If the signal and idler focus are
fixed, the optimal pump focus is given by $\xi_{\text{p}}=\xi_{\text{s}}%
(=\xi_{\text{i}})$. If instead $\xi_{\text{p}}$ is fixed, the optimal value of
$\xi_{\text{s}}$ is nearly equal to $\xi_{\text{p}}$ for $\xi_{\text{p}} \gtrsim 10$
and slightly larger than $\xi_{\text{p}}$ for $\xi_{\text{p}} \lesssim 10$.

\section{Spectral Purity}\label{Purity Sxn}

A number of applications of SPDC photons, including the
generation of multiphoton entangled states for quantum computing
\cite{Lu2007,Radmark2009,Wieczorek2009,Prevedel2009}, involve interference between photons from
separate (and nominally identical) SPDC sources. The success of such
applications depends not only on the efficiency of SPDC photon production, but
also on the degree of mutual coherence of photons from independent sources \cite{U'Ren2003,Humble2008}.
 This coherence, which is directly related to the interference visibility, is
given by the single-photon purity
\begin{equation}
\rho=\frac{\sum_{j}\sigma_{j}^{2}}{\left(  \sum_{j}\sigma_{j}\right)  ^{2}%
}\label{purity def}%
\end{equation}
where
\begin{equation}
\psi(\omega_{\text{s}},\omega_{\text{i}})=\sum_{j}\sqrt{\sigma_{j}}%
u_{j}(\omega_{\text{s}})v_{j}(\omega_{\text{i}})
\end{equation}
is the Schmidt decomposition \cite{Law2000} of the collected biphoton
state. The single-photon purity is inversely related to the degree of
entanglement between the signal and idler frequencies, with $\rho=1$
corresponding to no spectral entanglement. (Note that any spatial entanglement
present in the emitted state is discarded upon post-selecting photon pairs
that couple into the Gaussian collection modes.)

It is natural to ask whether the parameters that yield high source brightness
also yield high spectral purity. Under approximations (\ref{frequency
independence approx}) and (\ref{linear dispersion approx}), the
frequency-dependent part of $\psi(\omega_{\text{s}},\omega_{\text{i}})$ is
$s(\omega_{\text{p}})F(\xi,\Phi)$. Since the purity depends only on the
shape of $\psi(\omega_{\text{s}},\omega_{\text{i}}),$ and not on its location
or extent within the $(\omega_{\text{s}},\omega_{\text{i}})$ plane,
$\omega_{\text{s}}$ and $\omega_{\text{i}}$ may be replaced with the
dimensionless variables $\omega_{\text{s}}/\Omega$ and $\omega_{\text{i}%
}/\Omega$, where $\Omega\equiv c/(n_{\text{s}}^{\prime}-n_{\text{i}}^{\prime
})L$. Given a particular functional form for the pump spectrum, the purity may
then be computed as a function of $(\xi,\theta,\Delta\omega_{\text{p}}%
/\Omega)$ where $\Delta\omega$ is the pump bandwidth and
\begin{equation}
\tan \theta = \frac{n_{\text{s}}^{\prime}-n_{\text{p}}^{\prime}}{n_{\text{p}}^{\prime}-n_{\text{i}}^{\prime}}%
\label{theta def}
\end{equation}
is the slope, in the $(\omega_{\text{s}},\omega_{\text{i}})$ plane, of the line characterized by $\Phi=0$.
Note that since $\xi$ is the only relevant focusing parameter, we are free to set
$\xi_{\text{s}}\approx\xi_{\text{i}}\approx\xi_{\text{p}}$ to maximize the
brightness at a given $(\xi,\theta,\Delta\omega_{\text{p}}/\Omega)$. Also,
note that the brightness is independent of pump bandwidth when approximation
(\ref{linear dispersion approx}) is valid.

\begin{figure}
	\centering
		\includegraphics{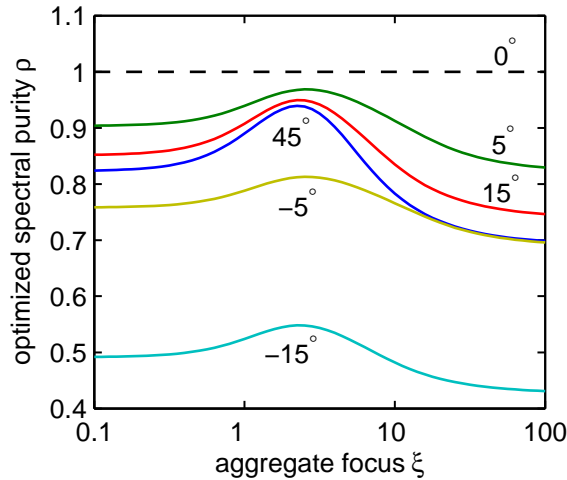}
	\caption{(color online) Dependence of the spectral purity of the collected photons on the mode focus. The numbers accompanying each line are the group velocity angle $\theta$ defined in eq.\ (\ref{theta def}).  Solutions for $\theta \ge 45^\circ$ mirror those for $\theta \le 45^\circ$.}
	\label{fig: spectral purity}
\end{figure}

The spectral purity obtainable with a Gaussian pump was determined by
computing $s(\omega_{\text{p}})F(\xi,\Phi)$ over a grid of sufficient extent
and resolution in the $(\omega_{\text{s}}/\Omega,\omega_{\text{i}}/\Omega)$
plane for various parameter values $(\xi,\theta,\Delta\omega_{\text{p}}%
/\Omega)$. For each set of parameter values, singular value decomposition was
applied to the matrix of computed values to determine $\{\sigma_{j}\}$ and
$\rho$. Numerical optimization was then used to determine the value of
$\Delta\omega_{\text{p}}/\Omega$ that maximizes $\rho$ at each $(\xi,\theta)$. In Fig.\ \ref{fig: spectral purity} these optimized values of $\rho$ are plotted as a function of
$\xi$ $(=\xi_{\text{p}})$ for several different values of $\theta$. For
$\theta=45^\circ$, the peak purity is 0.94 and occurs at $\xi=2.2$.  As $\theta$ decreases to $0^\circ$, $\rho$ asymptotically increases to 1 for all values of $\xi$, although the optimal pump bandwidth becomes infinite in this limit. Below $0^\circ$, $\rho$ drops rapidly.  Solutions for $\theta \ge 45^\circ$ mirror those for $\theta \le 45^\circ$, with the signal and idler exchanging roles.

%\begin{REVISED}
These results can be understood as follows.  The purity is directly related to the factorability of
$\psi(\omega_{\text{s}},\omega_{\text{i}}) \propto s(\omega_{\text{p}}) F(\xi,\Phi)$ into separate functions of $\omega_{\text{s}}$ and $\omega_{\text{i}}$.  When the pump function $s(\omega_{\text{p}})$ is Gaussian, complete factorability can be achieved if and only if the phase matching function $F(\xi,\Phi)$ is also Gaussian with orientation $0 \le \theta \le \pi/2$ \cite{Grice2001}. But $F(\xi,\Phi)$ is not Gaussian: for $\xi \ll 1$ it is a $\operatorname{sinc}$ function, which has oscillating side lobes, while for $\xi \gg 1$ it has large skew.  In both these regimes, the purity is reduced. $F(\xi,\Phi)$ is most nearly Gaussian for $\xi \sim 2$, which becomes the regime of greatest purity.
%\end{REVISED}

Comparing with Figs.\ \ref{fig: peak density image} and \ref{fig: pair flux line}, one sees that in the regime of peak purity ($\xi \sim 2$), the spectral
intensity is close to its peak value and the total collection probability is
more than 70\% of its asymptotic maximum (\ref{max pair flux}). Thus, high
spectral purity ($\geq94\%$) can be obtained with focusing conditions that
also yield relatively high brightness, provided $0\le\theta\le 90^\circ$ (which amounts to the requirement that $n_{\text{p}}^{\prime}$ lie between $n_{\text{s}}^{\prime}$ and $n_{\text{i}}^{\prime}$).

An explicit evaluation of the accuracy of these results in view of approximation
(\ref{C small approx}) was not performed. However, it may be noted that the
results of this section are based on a decomposition of an approximate state
which, as discussed at the end of section \ref{Bandwidth Sxn}, has a high
degree of overlap with the actual state.

\section{Single-Photon Collection and Heralding Ratio}\label{Herald Sxn}

In some applications of SPDC, the detection of a photon in
the signal mode is used to indicate the (probable) presence of a photon in the
idler mode.
%\begin{REVISED}
(Although the photons are always emitted in pairs, conditions generally allow
one photon to be emitted into a spatial mode defined by the collection optics
while its partner is emitted into a non-collected spatial mode.)
%\end{REVISED}
In such applications it is desirable to have a high
\emph{heralding ratio }$\eta_{\text{s}}\equiv P_{\text{si}}/P_{\text{s}},$
where $P_{\text{s}}$ is the probability that a photon is emitted into the
signal collection mode, regardless of the mode its partner is in.
%\begin{REVISED}
Since the probability of collecting both photons cannot exceed the probability
of collecting one of the photons, the heralding ratio can be at most unity.
%\end{REVISED}
In applications that treat signal and idler photons symmetrically, the quantity
$\eta_{\text{si}}\equiv P_{\text{si}}/\sqrt{P_{\text{s}}P_{\text{i}}}$ is
often the metric of choice. In this work $\eta_{\text{s}}$ will be called
the signal heralding ratio and $\eta_{\text{si}}$ will be called the symmetric
heralding ratio.

To obtain the signal collection probability $P_{\text{s}},$ the joint
collection probability must be summed over a complete set of spatial modes for
the idler photon. The Laguerre-Gauss modes
\begin{equation}
\mathbf{E}^{(n,l)}(\mathbf{r};\omega)=\frac{\mathbf{e}}{\sqrt{\pi/2}}\left(
\frac{w_{\text{i}}}{q_{\text{i}}}\right)  ^{l+1}\left(  \frac{q_{\text{i}%
}^{\ast}}{q_{\text{i}}}\right)  ^{n}L_{n}^{l}\left(  \frac{2w_{\text{i}}%
^{2}\rho^{2}}{\left|  q_{\text{i}}\right|  ^{2}}\right)  \exp\left[
-\frac{\rho^{2}}{q_{\text{i}}}+ik_{\text{i}}z+il\phi\right]
\end{equation}
form such a set, where $L_{n}^{l}$ is the $(n,l)$ associated Laguerre
polynomial, $\rho=\sqrt{x^{2}+y^{2}},$ and $\tan\phi=x/y$. The fundamental
mode with $n=l=0$ is just the Gaussian mode introduced in section \ref{Theory
Sxn}. Since the pump and signal mode are azimuthally symmetric, the spatial
overlap vanishes unless the idler mode is also azimuthally symmetric ($l=0$).
Thus the spatial overlap involving the $n$th idler mode is \
\begin{multline}
\mathcal{O}_{n} = \frac{\sqrt{\epsilon}\chi_{\text{eff}}^{(2)}}{\left(
\pi/2\right)  ^{3/2}}\int_{-L/2}^{L/2}\int_{0}^{\infty}\frac{w_{\text{p}%
}w_{\text{s}}w_{\text{i}}}{q_{\text{p}}q_{\text{s}}^{\ast}q_{\text{i}}^{\ast}%
}\left(  \frac{q_{\text{i}}}{q_{\text{i}}^{\ast}}\right)  ^{n}L_{n}\left(
\frac{2w_{\text{i}}^{2}\rho^{2}}{\left|  q_{\text{i}}\right|  ^{2}}\right)
\\
\times \exp\left[  -\rho^{2}\left(  \frac{1}{q_{\text{p}}}+\frac{1}{q_{\text{s}%
}^{\ast}}+\frac{1}{q_{\text{i}}^{\ast}}\right)  +i(\Delta k+mK)z\right]
2\pi\rho\,d\rho\,dz.
\end{multline}
With the Laguerre polynomial expansion formula $L_{n}(x)=\sum_{j=0}%
^{n}(-1)^{j}\frac{n!}{(n-j)!j!j!}x^{j}$ and a bit of work, one can obtain
\begin{equation}
\mathcal{O}_{n}=\sqrt{\frac{8\epsilon}{\pi}}\chi_{\text{eff}}^{(2)}%
w_{\text{p}}w_{\text{s}}w_{\text{i}}\int_{-L/2}^{L/2}\frac{\exp\left[
i(\Delta k+mK)z\right]  }{q_{\text{s}}^{\ast}q_{\text{i}}^{\ast}+q_{\text{p}%
}q_{\text{i}}^{\ast}+q_{\text{p}}q_{\text{s}}^{\ast}}\left(  \frac
{q_{\text{s}}^{\ast}q_{\text{i}}+q_{\text{p}}q_{\text{i}}-q_{\text{p}%
}q_{\text{s}}^{\ast}}{q_{\text{s}}^{\ast}q_{\text{i}}^{\ast}+q_{\text{p}%
}q_{\text{i}}^{\ast}+q_{\text{p}}q_{\text{s}}^{\ast}}\right)  ^{n}dz.
\end{equation}
Applying definitions (\ref{phi def})-(\ref{C def}) and making the
approximation $C\approx0$ yields
\begin{equation}
\mathcal{O}_{n}\approx\chi_{\text{eff}}^{(2)}\sqrt{\frac{2\epsilon}{\pi}%
\frac{k_{\text{s}}k_{\text{i}}}{k_{\text{p}}}L}\sqrt{\frac{\xi_{\text{s}}%
\xi_{\text{i}}}{\xi_{\text{p}}}}\int_{-1}^{1}\frac{\exp\left[  i\Phi
l/2\right]  }{A_{+}-ilB_{+}\xi_{\text{s}}\xi_{\text{i}}/\xi_{\text{p}}}\left(
\frac{A_{-}+ilB_{-}\xi_{\text{s}}\xi_{\text{i}}/\xi_{\text{p}}}{A_{+}%
-ilB_{+}\xi_{\text{s}}\xi_{\text{i}}/\xi_{\text{p}}}\right)  ^{n}dl.
\end{equation}

\begin{figure}
	\centering
		\includegraphics{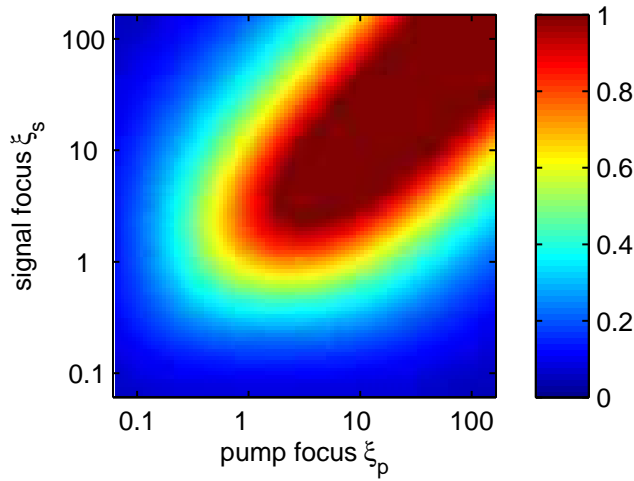}
	\caption{(color online) Dependence of the signal collection probability on the focus of the pump and signal modes for near-degenerate SPDC ($k_\text{s} \approx k_\text{i} \approx k_\text{p}/2$).}
	\label{fig: signal flux image}
\end{figure}

In analogy with formulas (\ref{flux def}) and (\ref{amplitude def}), the
signal photon probability may be written as $P_{\text{s}}=\int\sum
_{n=0}^{\infty}\left|  \psi_{n}(\omega_{\text{s}},\omega_{\text{i}})\right|
^{2}d\omega_{\text{s}}d\omega_{\text{i}}$ where $\psi_{n}=\sqrt{2\pi^{2}\hbar
N_{\text{p}}/\varepsilon_{0}\lambda_{\text{p}}\lambda_{\text{s}}%
\lambda_{\text{i}}}s(\omega_{\text{p}})\mathcal{O}_{n}(\omega_{\text{s}%
},\omega_{\text{i}})$. Following the approach taken in section \ref{Flux Sxn},
we have
\begin{align}
P_{\text{s}}  & \approx\frac{8\pi^{2}\hbar c\epsilon n_{\text{s}}n_{\text{i}%
}N_{\text{p}}}{\varepsilon_{0}n_{\text{p}}\left|  n_{\text{s}}^{\prime
}-n_{\text{i}}^{\prime}\right|  }\left(  \frac{\chi_{\text{eff}}^{(2)}%
}{\lambda_{\text{s}}\lambda_{\text{i}}}\right)  ^{2}\frac{\xi_{\text{s}}%
\xi_{\text{i}}}{\xi_{\text{p}}} \nonumber \\
& \hspace{2cm} \times \sum_{n=0}^{\infty} \int \left|  \int_{-1}^{1}%
\frac{\exp\left[  i\Phi l/2\right]  }{A_{+}-ilB_{+}\xi_{\text{s}}\xi
_{\text{i}}/\xi_{\text{p}}}\left(  \frac{A_{-}+ilB_{-}\xi_{\text{s}}%
\xi_{\text{i}}/\xi_{\text{p}}}{A_{+}-ilB_{+}\xi_{\text{s}}\xi_{\text{i}}%
/\xi_{\text{p}}}\right)  ^{n}dl\right|  ^{2}d\Phi\\
& =\frac{32\pi^{3}\hbar c\epsilon n_{\text{s}}n_{\text{i}}N_{\text{p}}%
}{\varepsilon_{0}n_{\text{p}}\left|  n_{\text{s}}^{\prime}-n_{\text{i}%
}^{\prime}\right|  }\left(  \frac{\chi_{\text{eff}}^{(2)}}{\lambda_{\text{s}%
}\lambda_{\text{i}}}\right)  ^{2}\xi_{\text{s}}\int_{-1}^{1}\frac
{dl}{A_{\text{s}}^{2}+(B_{\text{s}}\xi_{\text{s}})^{2}l^{2}}%
\end{align}
where
\begin{align}
A_{\text{s}}  & =2\sqrt{\left(  1+\frac{k_{\text{s}}}{k_{\text{p}}}\frac
{\xi_{\text{s}}}{\xi_{\text{p}}}\right)  \frac{k_{\text{i}}}{k_{\text{p}}}},\\
B_{\text{s}}  & =2\left(  1-\frac{\Delta k}{k_{\text{p}}}\right)
\sqrt{\left(  1+\frac{k_{\text{s}}+\Delta k}{k_{\text{p}}-\Delta k}\frac
{\xi_{\text{p}}}{\xi_{\text{s}}}\right)  \frac{k_{\text{i}}+\Delta
k}{k_{\text{p}}-\Delta k}}.
\end{align}
Integration yields the signal collection probability
\begin{equation}
P_{\text{s}}\approx\frac{64\pi^{3}\hbar c\epsilon n_{\text{s}}n_{\text{i}}%
}{\varepsilon_{0}n_{\text{p}}\left|  n_{\text{s}}^{\prime}-n_{\text{i}%
}^{\prime}\right|  }\left(  \frac{\chi_{\text{eff}}^{(2)}}{\lambda_{\text{s}%
}\lambda_{\text{i}}}\right)  ^{2}\frac{\arctan\left(  \frac{B_{\text{s}}%
}{A_{\text{s}}}\xi_{\text{s}}\right)  }{A_{\text{s}}B_{\text{s}}}N_{\text{p}%
}.\label{signal flux formula}%
\end{equation}
The corresponding formula for the idler probability $P_{\text{i}}$ can be
obtained by interchanging the labels s and i everywhere. Eq.\ (\ref{signal
flux formula}) is very similar to eq.\ (\ref{coinc flux formula}) and holds
under essentially the same conditions. Like $P_{\text{si}},$ the signal
probability $P_{\text{s}}$ is (to first approximation) an asymptotically
increasing function of focal parameters, with an upper bound
\begin{equation}
P_{\text{s}}\leq\frac{32\pi^{4}\hbar c\epsilon n_{\text{s}}n_{\text{i}}%
}{3\varepsilon_{0}n_{\text{p}}\left|  n_{\text{s}}^{\prime}-n_{\text{i}%
}^{\prime}\right|  }\left(  \frac{\chi_{\text{eff}}^{(2)}}{\lambda_{\text{s}%
}\lambda_{\text{i}}}\right)  ^{2}N_{\text{p}}.
\end{equation}
Also like $P_{\text{si}},$ $P_{\text{s}}$ is locally maximized by taking
$\xi_{\text{s}}\approx\xi_{\text{p}}$ (see Fig.\ \ref{fig: pair flux image}); however
$P_{\text{s}}$ varies more slowly with $\xi_\text{s}$ and $\xi_\text{p}$ than $P_{\text{si}}$ (Fig.\ \ref{fig: signal flux image}).
%\begin{REVISED}
The fact that $P_{\text{s}}$ is broader than $P_{\text{si}}$ means that collection of the signal in a non-optimal Gaussian mode projects the idler onto a mode that does not couple well to a Gaussian mode of any size, i.e. a mode that is not Gaussian.
%\end{REVISED}
Note also that (\ref{signal flux formula}) is independent of $\xi_{\text{i}}$ as it should be (the
parameters of the idler collection mode should be irrelevant when one does not
care whether the idler is collected).

\begin{figure}
	\centering
		\includegraphics{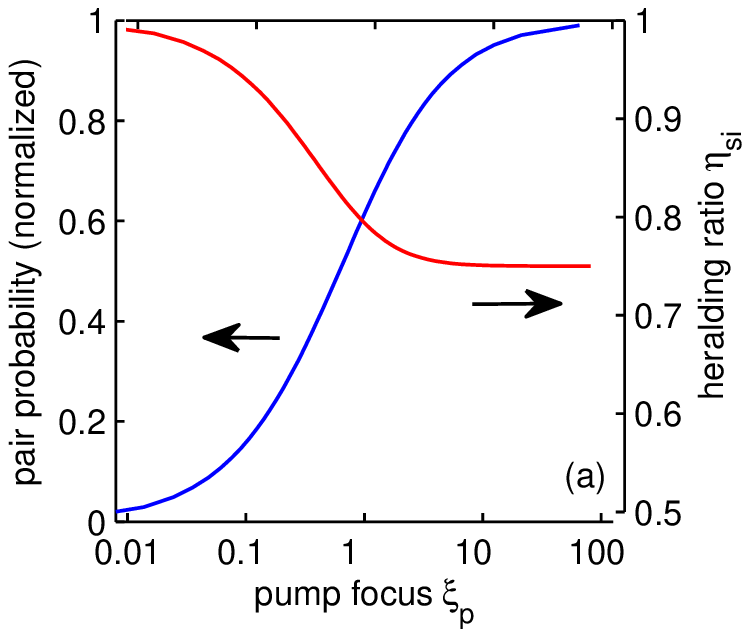}
		\includegraphics{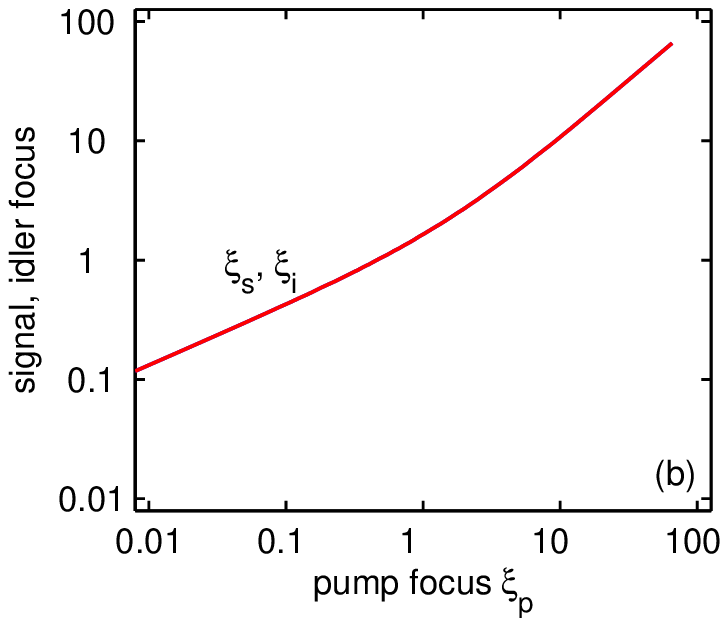} \\
		\includegraphics{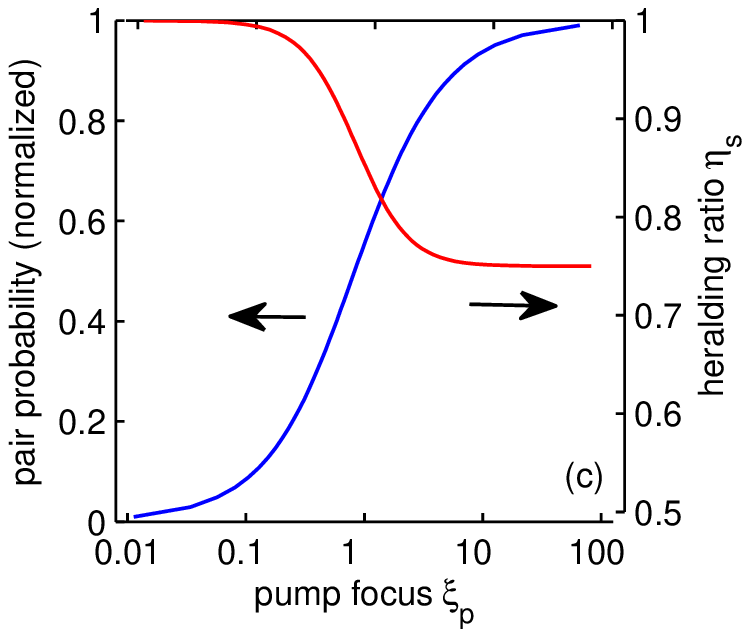} 
		\includegraphics{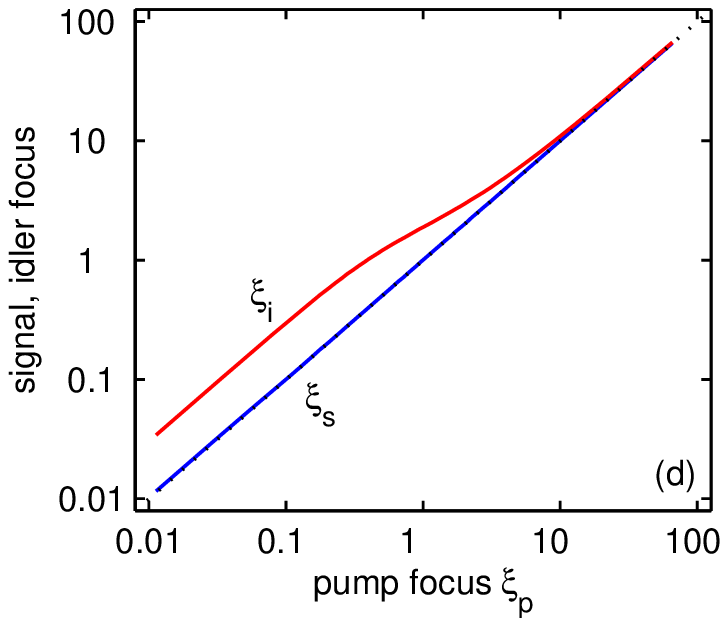}
	\caption{(color online) (\emph{a,b}) Simultaneous optimization of the total collection probability $P_\text{si}$ and the symmetric heralding ratio $\eta_\text{si}$. (\emph{c,d}) Simultaneous optimization of $P_\text{si}$ and the signal heralding ratio $\eta_\text{s}$. Panels (\emph{b}) and (\emph{d}) plot the focal parameters that yield the optimal curves shown in (\emph{a}) and (\emph{c}), respectively.  In (\emph{a}) and (\emph{c}) the collection probability goes with the left axis and the heralding ratio goes with the right axis.}
	\label{fig: heralding}
\end{figure}

If one takes $\xi_{\text{s}}=\xi_{\text{i}}=\xi_{\text{p}},$ which locally
maximizes both $P_{\text{si}}$ and $P_{\text{s}},$ the signal heralding ratio
reduces to the simple expression
\begin{equation}
\eta_{\text{s}}=\frac{k_{\text{i}}}{k_{\text{p}}}\left(  \frac{k_{\text{s}}%
}{k_{\text{p}}}+1\right)  .
\end{equation}
An analogous expression exists for $\eta_{\text{i}}$. These expressions give
$\eta_{\text{s}}=\eta_{\text{i}}=0.75$ for near-degenerate SPDC (for which
$k_{\text{s}}\approx k_{\text{i}}\approx k_{\text{p}}/2$) and values less than
0.75 for the non-degenerate case. However, higher heralding ratios are
possible with different focusing conditions. The optimal source
configuration with regard to both heralding and brightness is not a single set
of parameter values, but a curve in parameter space having the property that
$\eta_{\text{s}}$ (or $\eta_{\text{si}},$ if that is of interest) cannot be
increased without decreasing $P_{\text{si}},$ and vice versa. Numerical
methods were used to find the values $(\xi_{\text{s}},\xi_{\text{i}}%
,\xi_{\text{p}})$ that maximize either $\eta_{\text{s}}$ or $\eta_{\text{si}}$
at 50 values of $P_{\text{si}}$ covering the range from 0 to $\max
P_{\text{si}}$. The resulting points are plotted versus $\xi_{\text{p}}$ in
Fig.\ \ref{fig: heralding} for the case $k_{\text{s}}\approx k_{\text{i}}\approx k_{\text{p}%
}/2$. Fig.\ \ref{fig: heralding} shows that it is possible to achieve very high heralding
ratios, but only with a substantial reduction in brightness: a factor of at
least 4 to achieve $\eta_{\text{s}}\geq0.95$ and a factor of at least 10 to
achieve $\eta_{\text{si}}\geq0.95$. That is, there is a trade-off between
brightness and heralding ratio, which is somehwat worse for symmetric heralding
than asymmetric heralding. Both $\eta_{\text{s}}$ and $\eta_{\text{si}}$
approach unity in the limit that the pump is collimated ($\xi_{\text{p}%
}\rightarrow0$). In this limit the best trade-off between $P_{\text{si}}$
and $\eta_{\text{si}}$ is achieved with $\xi_{\text{s}}=\xi_{\text{i}}\gg
\xi_{\text{p}},$ while the best trade-off between $P_{\text{si}}$ and
$\eta_{\text{s}}$ is achieved with $\xi_{\text{i}}\approx3\xi_{\text{s}%
}\approx3\xi_{\text{p}}$. The trade-offs between $P_{\text{si}}$ and
$\eta_{\text{s}}$ or $\eta_{\text{si}}$ are found to be slightly worse in the
non-degenerate case than the near-degenerate case.

\section{Discussion}\label{Discussion Sxn}

Perhaps the most significant finding of this study is
that there exist focusing conditions ($\xi_{\text{p}},\xi_{\text{s}}%
,\xi_{\text{i}}\sim2.5$) that simultaneously bring the brightness, heralding
ratio, and spectral purity to substantial fractions of their maximal values.
Whenever $n_{\text{p}}^{\prime}$ lies between $n_{\text{s}}^{\prime}$ and
$n_{\text{i}}^{\prime},$ a spectral purity of at least 0.94 can be obtained
with at least 74\% of the maximum achievable brightness, while the heralding
ratio is at least 0.75 (for near-degenerate SPDC). Maximum brightness can be
achieved with tighter focusing at the cost of reducing the spectral purity (to
0.7 at worst) The worst trade-off is between brightness and heralding ratio:
focusing conditions that yield a symmetric (asymmetric) heralding ratio of
0.95 or better reduce the brightness to 10\% (25\%) of its maximal value;
however the spectral purity remains 0.84 or better.

Another notable finding of this study, which to my knowledge has not been
reported previously, is that many important properties of the collected SPDC
state are essentially scale invariant. To the extent that the parameter $C$
(see (\ref{C def})) is negligibly small, the joint collection probability,
heralding ratios, and spectral purity are determined by the dimensionless
ratios $L/k_{j}w_{j}^{2}$ and $\Delta\omega_{\text{p}}L/c$. Changing the
crystal length has no effect on these three quantities if the pump duration
and confocal ranges are also scaled by the same factor. In the end, the
crystal length merely sets the bandwidth of the system, with longer crystals
yielding (or requiring) smaller bandwidths. The idea that a longer crystal
does not yield a brighter source is perhaps surprising, since longer crystals
yield higher SPDC efficiencies when the pump is collimated or weakly focused.
Why this is not so for \emph{optimally focused} sources may be understood by
noting that, although a shorter crystal has a shorter interaction length
(which decreases the spatial mode overlap), it also allows the modes to be
focused more tightly (which increases the mode overlap). Of course, this
argument does not hold for arbitrarily short crystals; at some point, the
focusing becomes so strong and the bandwidths become so large that the
paraxial approximation implicit in (\ref{Gaussian mode}), as well as
approximations such as (\ref{frequency independence approx}), break down. At
the other extreme, very long crystals may also show worse than expected
performance due to the challenge of manufacturing very long crystals of high quality.

Some of the results obtained here appear to agree with prior works, while
others appear to differ from prior works. The focusing condition
$\xi_{\text{s}}=\xi_{\text{i}}=\xi_{\text{p}}=2.84$ which maximizes the joint
spectral density of SPDC is the same as that found by Boyd and Kleinman for
maximizing second harmonic generation \cite{Boyd1968}. This makes sense,
since SHG with a monochromatic Gaussian input beam produces a monochromatic
Gaussian second harmonic field; optimization of SHG then amounts to finding
the parameters of the Gaussian modes that maximize the spatial overlap, which
is precisely what was done in section \ref{Intensity Sxn}. More recently,
Ljunggren and Tengner \cite{Ljunggren2005} have performed numerical studies yielding
the optimizing conditions $2\xi_{\text{p}}=1.7,$ $2\xi_{\text{s}}%
=2\xi_{\text{i}}=2.3$, and the prediction that the optimized joint probability
goes as $\sqrt{L}$ (rather than being independent of $L$ as claimed here).
%In \cite{Kolenderski2009}, analytical and numerical calculations show the condition
%$w_\text{s}=w_\text{i}=2w_\text{p}$ to be optimal (amounting to $\xi_\text{s} = \xi_\text{i}
% = \xi_\text{p}/2$ in the frequency-degenerate case), but that result is for non-collinear beams.
%The discrepencies between this study and \cite{Ljunggren2005}
These discrepencies may be due to two key differences
in approach: In \cite{Ljunggren2005}, optimal focusing conditions are obtained for $\Phi=0$,
whereas here and in \cite{Boyd1968} the condition $\Phi=-1.04\pi$ is found to
yield a slightly higher spectral density. Additionally, \cite{Ljunggren2005}
employs a plane-wave approach in which the diffraction of the pump mode is
effectively ignored; this approximation is also made in \cite{Castelletto2005},
\cite{Dragan2004} and \cite{Ling2008}. In the present study the pump is treated as a
diffracting paraxial beam. Experiments to date are insufficient to resolve
these differences as they cover a limited range of crystal lengths and focal
parameters, and are complicated by poor repeatability (changing the focus of a
beam, or replacing a crystal with one of different length, generally
necessitates realignment).

The results of this study are intended to be useful for designing and
predicting the approximate performance of a promising class of SPDC sources.
 A number of points should be kept in mind, however. Firstly, all the
results presented here apply to the \emph{collected} part of the biphoton
state; many features of the emitted SPDC light are lost or substantially
altered when the state is projected onto the Gaussian collection modes. Secondly, while care
was taken to confirm the general viability of the approximations employed, it
is not difficult to envision sources that would violate the assumptions of
this analysis and exhibit quantitatively or qualitatively different
performance. In particular, results obtained here may not be reliable for
very short sources ($L \lesssim \operatorname{mm}$), sources with very short
poling periods ($\Lambda \lesssim 5 \operatorname{\mu m}$),
and/or sources employing quasi-phasematching of order $m>1$. Thirdly,
certain physical details deemed inconsequential---such as the vector
diffraction of the Gaussian beams, the optical anisotropy of nonlinear
crystals, and the proper normalization of electromagnetic modes quantized in a
dielectric---were simply ignored. Fourthly, a limitiation of this study is
that it does not apply to angle-tuned sources in which beam walk-off plays a
role. However, since walk-off decreases spatial overlap, it is not clear
that such sources could achieve better overall performance than the co-linear
sources considered here. Finally, of all the results obtained here, only
those in section \ref{Intensity Sxn} are potentially applicable to
frequency-degenerate type I sources. The quadratic relationship between
frequency and phase mismatch in these sources substantially complicates
analysis; however, the possibility that these sources might have different scaling laws makes them
potentially worth a comparable study.

\section{Summary}\label{Summary Sxn}
In summary, I have presented here a new theoretical study of SPDC addressing
multiple properties of the emitted photons that are important in various
applications. Analysis was restricted to the promising class of SPDC sources
involving focused, co-linear Gaussian modes for the pump field and collected
photons. Analytical and numerical calculations yielded approximate predictions
for the peak spectral density, photon bandwidths, absolute pair collection
probability, heralding ratio, and spectral purity. A scaling law was found
which shows most of these properties to be ultimately independent of crystal
length. It was also found that such sources can simultaneously exhibit
high brightness (predicted $10^{-9}$ pairs/pump photon), high spectral
purity ($\geq0.94$), and moderately high heralding ratio ($\gtrsim0.75$) when
the confocal ranges of the modes are on the order of half the crystal length.
Higher heralding ratios can be achieved, at the cost of significantly reduced
brightness, by focusing the modes less tightly.  The results of this study are
applicable to ``typical'' SPDC sources, excluding sources of very short
length, very short poling period, or very large bandwidth.  Frequency-degenerate type
I sources, which often do not satisfy these criteria, may be amenable to a similar kind of study.

This work was sponsored by the Intelligence Advanced Research Projects Activity (IARPA)
and by the Laboratory Directed Research and Development Program
of Oak Ridge National Laboratory, managed by UT-Battelle, LLC, for the U. S. Department
of Energy.

The author thanks W. P. Grice, T. S. Humble, and P. G. Evans for helpful discussions.

%\bibliographystyle{ryan-unsrt}
%\bibliography{QuantumOptics}

\begin{thebibliography}{10}
\newcommand{\enquote}[1]{``#1''}

\bibitem{Politi2009_IEEE}
A.~Politi, J.~C.~F. Matthews, M.~G. Thompson, and J.~L. O'Brien.
\newblock \emph{IEEE Journal on Selected Topics in Quantum Electronics}
  \textbf{15}, 1673 (2009).

\bibitem{SECOQC2009}
M.~Peev et al.
\newblock \emph{New Journal of Physics} \textbf{11}, 075001 (2009).

\bibitem{Wong2006}
F.~Wong, J.~Shapiro, and T.~Kim.
\newblock \emph{Laser Physics} \textbf{16}, 1517 (2006).

\bibitem{Fedrizzi2007}
A.~Fedrizzi, T.~Herbst, A.~Poppe, T.~Jennewein, and A.~Zeilinger.
\newblock \emph{Optics Express} \textbf{15}, 598  (2007).

\bibitem{Grice2001}
W.~P. Grice, A.~B. U'Ren, and I.~A. Walmsley.
\newblock \emph{Phys. Rev. A} \textbf{64}, 063815 (2001).

\bibitem{U'Ren2007}
A.~U'Ren, Y.~Jeronimo-Moreno, and H.~Garcia-Gracia.
\newblock \emph{Physical Review A}
  \textbf{75}, 23810 (2007).

\bibitem{Mosley2009}
P.~J. Mosley, J.~S. Lundeen, B.~J. Smith, and I.~A. Walmsley.
\newblock \emph{Journal of Modern Optics} \textbf{56}, 179  (2009).

\bibitem{Humble2007}
T.~Humble and W.~Grice.
\newblock \emph{Physical Review A}
  \textbf{75}, 022307 (2007).

\bibitem{Kurtsiefer2001}
C.~Kurtsiefer, M.~Oberparleiter, and H.~Weinfurter.
\newblock \emph{Phys. Rev. A} \textbf{64}, 023802  (2001).

\bibitem{Bovino2003}
F.~Bovino, P.~Varisco, A.~Colla, G.~Castagnoli, G.~Di~Giuseppe, and
  A.~Sergienko.
\newblock \emph{Opt. Comm.} \textbf{227}, 343  (2003).

\bibitem{Dragan2004}
A.~Dragan.
\newblock \emph{Phys. Rev. A} \textbf{70}, 053814 (2004).

\bibitem{Andrews2004}
R.~Andrews, E.~Pike, and S.~Sarkar.
\newblock \emph{Opt. Expr.} \textbf{12}, 3264 (2004).

\bibitem{Castelletto2005}
S.~Castelletto, I.~Degiovanni, G.~Furno, V.~Schettini, A.~Migdall, and M.~Ware.
\newblock \emph{IEEE Trans. Instrum. Meas.} \textbf{54}, 890  (2005).

\bibitem{Ljunggren2005}
D.~Ljunggren and M.~Tengner.
\newblock \emph{Phys. Rev. A} \textbf{72}, 062301  (2005).

\bibitem{Ling2008}
A.~Ling, A.~Lamas-Linares, and C.~Kurtsiefer.
\newblock \emph{Physical Review A}
  \textbf{77}, 043834 (2008).

\bibitem{Kolenderski2009}
P.~Kolenderski, W.~Wasilewski, and K.~Banaszek.
\newblock \emph{Physical Review A}
  \textbf{80}, 013811 (2009).

\bibitem{Hong1985_PRA}
C.~Hong and L.~Mandel.
\newblock \emph{Physical Review A} \textbf{31}, 2409 (1985).

\bibitem{ORNL_source} P.~Evans, J.~Schaake, R.~Bennink, T.~Humble, and W.~Grice (unpublished).

\bibitem{Boyd1968}
G.~D. Boyd and D.~A. Kleinman.
\newblock \emph{Journal of Applied Physics} \textbf{39}, 3597 (1968).

\bibitem{Carrasco2006}
S.~Carrasco, A.~V.~Sergienko, B.~E.~Saleh, M.~C.~Teich, J.~P.~Torres, and L.~Torner.
\newblock \emph{Phys. Rev. A} \textbf{73} 063802 (2006).

\bibitem{Kwiat1995}
P.~G.~Kwiat, K.~Mattle, H.~Weinfurter, A.~Zeilinger, A.~V.~Sergienko, and Y.~H.~Shih.
\newblock \emph{Phys. Rev. Lett.} \textbf{75}, 4337 (1995).

\bibitem{Zhao2008}
Z.~Zhao, K.~A.~Meyer, W.~B.~Whitten, R.~W.~Shaw, R.~S.~Bennink, and W.~P.~Grice.
\newblock \emph{Phys. Rev. A} \textbf{77}, 063828 (2008).

\bibitem{Lu2007}
C.-Y. Lu, X.-Q. Zhou, O.~Guhne, W.-B. Gao, J.~Zhang, Z.-S. Yuan, A.~Goebel,
  T.~Yang, and J.-W. Pan.
\newblock \emph{Nat Phys} \textbf{3}, 91 (2007).

\bibitem{Radmark2009}
M.~Radmark, M.~Zukowski, and M.~Bourennane.
\newblock \emph{Physical Review Letters} \textbf{103}, 150501 (2009).

\bibitem{Wieczorek2009}
W.~Wieczorek, R.~Krischek, N.~Kiesel, P.~Michelberger, G.~Toth, and
  H.~Weinfurter.
\newblock \emph{Physical Review Letters} \textbf{103}, 020504 (2009).

\bibitem{Prevedel2009}
R.~Prevedel, G.~Cronenberg, M.~Tame, M.~Paternostro, P.~Walther, M.~Kim, and
  A.~Zeilinger.
\newblock \emph{Physical Review Letters} \textbf{103}, 020503 (2009).

\bibitem{U'Ren2003}
A.~U'Ren, E.~Mukamel, K.~Banaszek, and I.~Walmsley.
\newblock \emph{Phil. Trans. Roy. Soc. Lond. A} \textbf{361}, 1493
  (2003).

\bibitem{Humble2008}
T.~Humble and W.~Grice.
\newblock \emph{Physical Review A}
  \textbf{77}, 022312 (2008).

\bibitem{Law2000}
C.~K. Law, I.~A. Walmsley, and J.~H. Eberly.
\newblock \emph{Phys. Rev. Lett.} \textbf{84}, 5304 (2000).

\end{thebibliography}

\end{document}